\documentclass[conference]{IEEEtran}
\IEEEoverridecommandlockouts
\usepackage{cite}
\usepackage{amsmath,amssymb,amsfonts}
\usepackage{algorithmic}
\usepackage{graphicx}
\usepackage{textcomp}
\usepackage{xcolor}
\def\BibTeX{{\rm B\kern-.05em{\sc i\kern-.025em b}\kern-.08em
    T\kern-.1667em\lower.7ex\hbox{E}\kern-.125emX}}
\begin{document}

\title{Energy Efficient End-to-End MIMO Wireless Autoencoder based on Spatio-Temporal Projection
\thanks{The work of Rodney Martinez Alonso is supported by a Senior Postdoctoral fellowship from the Research Foundation–Flanders (FWO) in the framework of the AIxG project under Grant 1211926N. This work was also supported under the 6G-MIRAI project, which has received funding from the Smart Networks and Services Joint Undertaking (SNS JU) under the European Union’s Horizon Europe research and innovation program under Grant Agreement No 101192369.}
}

\author{\IEEEauthorblockN{1\textsuperscript{st} Rodney Martinez Alonso}
	\IEEEauthorblockA{
		\textit{ESAT, KU Leuven}\\
		Leuven, Belgium \\
		0000-0003-2529-5944}
	\and
	\IEEEauthorblockN{2\textsuperscript{nd} Cel Thys}
	\IEEEauthorblockA{\textit{ESAT, KU Leuven}\\
		Leuven, Belgium \\
		0000-0002-3752-9456}
	\and
	\IEEEauthorblockN{3\textsuperscript{rd} Robbert Beerten}
	\IEEEauthorblockA{\textit{ESAT, KU Leuven}\\
		Leuven, Belgium \\
		0000-0002-2194-9202}
	\and
	\IEEEauthorblockN{4\textsuperscript{th} Emiel Vanspranghels       }
	\IEEEauthorblockA{\textit{ESAT, KU Leuven}\\
		Leuven, Belgium \\
		0009-0008-3086-0171}\\ 
	\and
	\IEEEauthorblockN{5\textsuperscript{th} George Mystriotis}
	\IEEEauthorblockA{\textit{ESAT, KU Leuven}\\
		Leuven, Belgium \\
		0009-0009-2455-8475}
	\and
	
	\IEEEauthorblockN{6\textsuperscript{th} Wout Joseph}
	\IEEEauthorblockA{\textit{INTEC, Ghent University -- imec}\\
		Ghent, Belgium \\
		0000-0002-8807-0673}
	\and
	\IEEEauthorblockN{7\textsuperscript{th} David Plets}
	\IEEEauthorblockA{\textit{INTEC, Ghent University -- imec}\\
		Ghent, Belgium \\
		0000-0002-8879-5076}
	\and
	\IEEEauthorblockN{8\textsuperscript{th} Sofie Pollin}
	\IEEEauthorblockA{\textit{ESAT, KU Leuven}\\
		Leuven, Belgium \\
		0000-0002-1470-2076}
}
\maketitle

\begin{abstract}
End-to-end wireless autoencoders have emerged as a promising method for AI-native wireless interfaces. However, a major challenge is their scalability and energy efficiency for spatially multiplexed MIMO communications. In this paper, we propose the design of a neural wireless autoencoder based on spatio-temporal expanded projection that decouples spatial and temporal processing. By decoupling the spatial and temporal domain processing, the neural receiver achieves better performance for a multiplexed channel with implicit cross-antenna correlation caused by the interleaving of the encoded bitwords. For example, for 4x4 MIMO and a 16-symbol non-uniform constellation, the model achieves an SNR gain of 2.2~dB compared to the equivalent dense neural-based autoencoder with in-quadrature modulation. The proposed model also improves the computational energy efficiency on an edge-grade accelerator by a factor of up to 4.6.
\end{abstract}

\begin{IEEEkeywords}
Wireless Autoencoder, Spatial Multiplexed MIMO, Neural Networks
\end{IEEEkeywords}

\section{Introduction}
The reconfigurability and scalability of the Physical Layer (PHY) have become a challenge for new wireless standards, as the PHY plays a key role in the tradeoff between energy efficiency and hardware life cycle. To address the fundamental gap behind suboptimal hardware adaptability as a function of energy efficiency in the PHY, different authors have proposed the design of an AI-native adaptable wireless interface~\cite{hoydis_2021_6gainative}. The primary objective of an AI-native air (wireless) interface is to customize the PHY as a function of the specific radio propagation environment and the hardware constraints~\cite{hoydis_2021_6gainative}. However, the energy efficiency of current neural models for the PHY layer in general, and specifically for spatially multiplexed Multi-Input Multi-Output (MIMO) channels, remains limited. Moreover, even in perfectly uncorrelated MIMO channels, the multiplexing of the same encoded bitwords across antennas creates implicit correlation.

One of the key technological enablers for the envisioned AI-Native reconfigurable PHY are end-to-end neural-based wireless autoencoders~(WAE)~\cite{dorner_learning_2023,ait_aoudia_waveform_2022,hesham_coding_2023,HybridE2E_2026}. The WAE use neural networks to jointly learn to encode and decode messages into a latent representation that passes through the physical communication channel (and analog hardware components of the transceivers)~\cite{Korpi_ICC_2026}. Unlike conventional block-based digital signal processing that defines and optimizes each functional block in isolation, the WAE jointly trains a neural-based transmitter and receiver as an integrated end-to-end system. In this way, the PHY can achieve task-oriented, channel, and hardware-aware performance optimization~\cite{TaskOriented_Comms,hoydis_2021_6gainative}.

Most of the work on AI-based PHY has focused on self-design for a single spatial channel. Only recently, major efforts have been made to extend the WAE to MIMO. For instance,~\cite{Velloso_2xMIMO_2026} present a MIMO autoencoder for 2 spatially multiplexed streams enabling Non-Uniform Constellations~(NUC). The design is based on a Dense Neural Network (DNN), including channel estimation and equalization. Despite the fact that the authors can decrease the pilot overhead and add constellation shaping gain, the system requires several heuristic iterations during inference that could not scale towards larger MIMO configurations. In~\cite{Song_2xMIMO_2022}, a 2x2 multiplexed MIMO autoencoder also based on DNN is investigated. In this approach, the symbols are encoded as a one-hot vector, which the neural network processes as a unique parametrized feature. This feature fingerprinting creates a critical bottleneck in scalability and growing complexity as the number of antennas and rate increase~\cite{Song_2xMIMO_2022}. The common research gaps in the literature are: 1) the lack of a neural-based wireless PHY addressing the computational and energy limitations at the edge; 2) scalable and energy-efficient MIMO autoencoding enabling the added gains from NUC shaping

The leap in knowledge from our research lies in the design of a neural-based MIMO autoencoder that applies low-complexity, parameterized NUC shaping in the user equipment (UE) across multiple Uplink streams, while asymmetrically shifting the primary computational burden to the access point. This is realized through a hybrid architecture that combines tractable digital signal processing blocks with neural-based NUC decoding to mitigate residual multiplexed MIMO interference. Furthermore, in the receiver, we separate the spatial and temporal processing to significantly maximize parallelization and computational energy efficiency on edge accelerators. This architecture also mitigates the impact of inter-symbol correlation caused by the interleaving of the encoded word across antennas.

\section{Hybrid-Asymmetric MIMO Autoencoder}
\label{sec:Method}
\begin{figure}[h]
	\centering
	\includegraphics[width=0.9\linewidth]{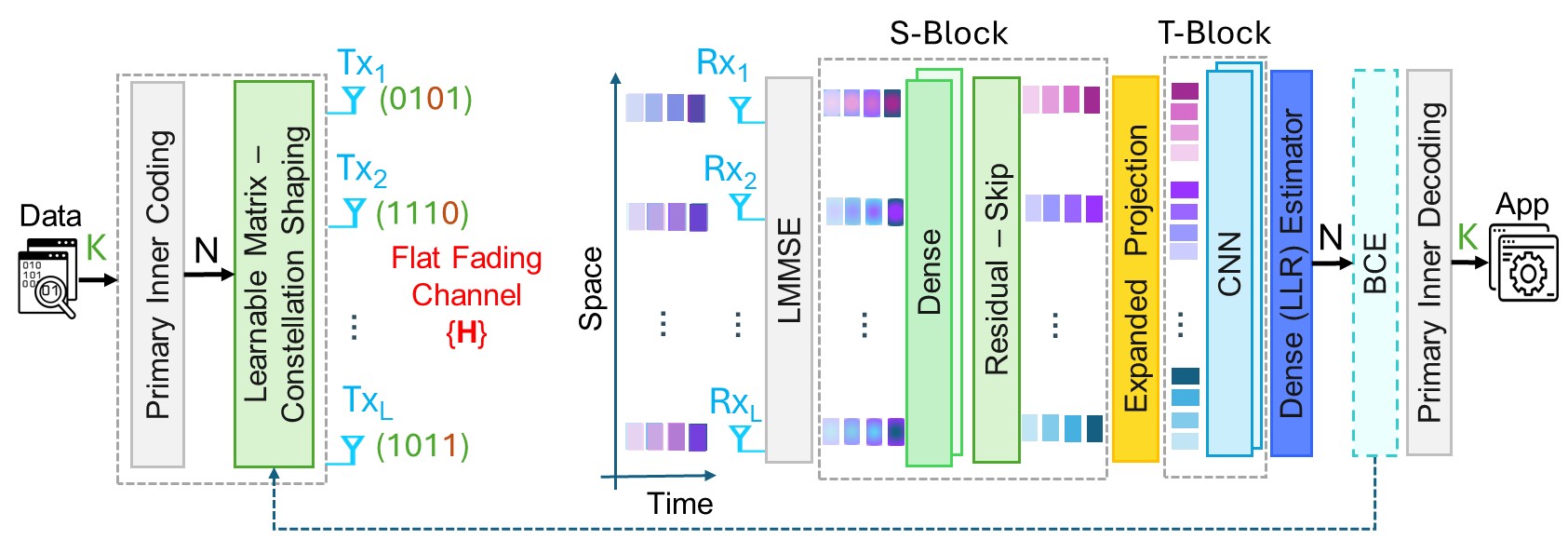}
	\caption{Overview of the Hybrid Asymmetric MIMO Wireless Autoencoder. S-Block: Spatial processing; T-Block: time-dimension processing after projection.}
	\label{fig:Hybrid_WAE_Diagram}
\end{figure}

A fundamental gap that we aim to tackle is the reduced computational resources for implementing a virtualized PHY on general-purpose constrained hardware. Therefore, for the uplink, our goal is to minimize the computational effort required by the transmitting UE (i.e., \textit{asymmetric} compute). Fig.~\ref{fig:Hybrid_WAE_Diagram} shows the proposed spatially multiplexed MIMO autoencoder with an asymmetric distribution of the computational complexity between the UE and the access point.

In this design, we mixed tractable and neural-based blocks (\textit{hybrid} wireless autoencoder). All the blocks were implemented using Sionna-PHY~\cite{sionna}. Most of the design is standard-agnostic. The main exception is the Primary Inner Coding for the forward error correction. We used the LDPC function defined in the Sionna-PHY~\cite{sionna} simulator, which follows the general guidelines from ETSI 3GPP TS 38.212~\cite{etsi3gpp38212}. This block encodes \textit{K} information bits into an \textit{N}-bit codeword. Explicit internal interleaving is omitted to bypass the predefined 5G modulation and coding schemes, as it constrains the possible combinations of $N,K$. Instead, the constellation-shaped symbols of size $Q_m$ are interleaved across the UE antennas (cf. Tx in Fig.~\ref{fig:Hybrid_WAE_Diagram}).

To reduce the transmitter complexity, we implement NUC shaping by directly optimizing a learnable matrix of the complex constellation points ($M=2^{Q_m}$) through backpropagation, rather than generating them with a neural network. The constellation is normalized to zero mean and unit average energy before symbol mapping. The index of the learnable matrix implicitly creates a learnable bit-label-to-symbol association. By directly optimizing this learnable matrix rather than utilizing multi-layer neural networks for symbol generation, the transmitter achieves geometric shaping capabilities with minimal parametric footprint (i.e., $2 \cdot M$ real parameters). The symbols are further interleaved across the $N_{tx}$ antennas of the transmitter.

We consider a flat-fading MIMO channel with ($N_{tx}$) transmit and ($N_{rx}$) receive antennas. The received signal is modeled as $\mathbf y=\mathbf H\mathbf x+\boldsymbol\eta$, where ($\mathbf x\in\mathbb C^{N_{tx}}$) is the transmitted symbol vector, ($\mathbf H\in\mathbb C^{N_{rx}\times N_{tx}}$) is the channel matrix, and $\boldsymbol\eta\sim\mathcal{CN}(\mathbf 0,\sigma_\eta^2\mathbf I_{N_{rx}})$ represents spatially uncorrelated additive Gaussian noise for a given noise variance ($\sigma_\eta^2$).

\subsection{LMMSE Equalization}

The receiver will apply a Linear Minimum Mean Squared Error (LMMSE) equalizer as a first low-compute-cost cross-antenna interference cancellation~\cite{EqDeepRx_2026}. This equalizer will compute a filter matrix $\mathbf{W}_{\text{LMMSE}} \in \mathbb{C}^{N_{tx} \times N_{rx}}$ that minimizes the mean squared error $\mathbb{E}\left[ \|\mathbf{x} - \hat{\mathbf{x}}\|^2 \right]$ between the transmitted symbols $\mathbf{x}$ and their linear estimates $\hat{\mathbf{x}}$. Assuming unit-variance transmit symbols ($\mathbb{E}[\mathbf{x}\mathbf{x}^H] = \mathbf{I}_{N_{tx}}$), the LMMSE weight matrix is derived as:
\begin{equation}
	\mathbf{W}_{\text{LMMSE}} = \mathbf{H}^H \left( \mathbf{H}\mathbf{H}^H + \mathbf{S} \right)^{-1},
\end{equation}
where $(\cdot)^H$ denotes the Hermitian transpose operator. Applying the equalization filter to $\mathbf{y}$ yields the estimated symbol vector $\hat{\mathbf{x}}$. To enable soft-demapping, the post-equalization effective noise variance for the $i$-th spatial layer can be computed from the diagonal elements of the error covariance matrix.

\subsection{Neural Decoding}

After the initial interference cancellation by LMMSE, we further use a neural receiver to jointly remove residual noise and perform the demapping (in Fig.~\ref{fig:Hybrid_WAE_Diagram}). At the output of the LMMSE block, the post-equalized symbol estimates $\hat{\mathbf{x}}$ and the spatial layer-specific effective noise variances $\boldsymbol{\sigma}_{\text{eff}}^2 \in \mathbb{R}^{N_{tx}}$ are subsequently provided to a neural receiver (cf. Fig.~\ref{fig:Hybrid_WAE_Diagram}). Therefore, the input tensor for the neural decoding stage is the complex symbol estimate $\hat{x}_{L,t}$ comprising the real and imaginary parts and concatenated with the noise variance as: 
\begin{equation}
	\mathbf{x}_{L,t} =
	\left[
	\Re(\hat{x}_{L,t}),
	\Im(\hat{x}_{L,t}),
	\sigma^2_{\text{eff},L,t}
	\right]^T
	\in \mathbb{R}^3
\end{equation}
where for $L \in \{1, \dots, N_{\text{t}}\}$ represent the number of separated spatial streams at temporal index $t \in \{1, \dots, N_{\text{sym}}\}$ obtained after LMMSE processing, and  $\sigma^2_{\text{eff}, L, t} \in \mathbb{R}^+$ denote the corresponding effective residual noise and interference variance. These vectors across all spatial streams and time slots form the feature tensor to input to the neural decoder:
\begin{equation}
	\mathbf{X} \in \mathbb{R}^{B \times L \times N_{\text{sym}} \times 3}
\end{equation}
where $B$ denotes the batch size, and the 3 dimensions refer to the I, Q, and noise components in the tensor from the previous block. The neural decoder aims to compute log-likelihood ratios $LLR \in \mathbb{R}^{B \times L \times N_{\text{sym}} \times M}$ for $M$ coded bits per symbol per spatial stream while effectively mitigating residual multi-user interference and tracking temporal channel variations without cross-stream data leakage. 

In previous DNN-based MIMO autoencoders, the vector's temporal dimension is not processed separately but is treated as part of the batch. This means the input dense layer size will scale as $d_{\text{in}} \propto \alpha \cdot \max(L, M)$, with $\alpha \geq 2$ in practice. Our design improves the performance and complexity tradeoff for the joint NUC shaping, cross-antenna noise cancellation, and demapping by projecting the signal vector across its temporal domain.  This approach will help further optimize the constellation shaping and add robustness against induced correlation from the interleaving of the same LDPC-encoded bitwords across antennas. The dense input layer will scale mainly with the number of antennas only (i.e., $d_{in} \propto  \alpha \cdot L$. These complexity boundaries are true as long as the combined amount of learnable parameters is enough to separate the constellation points.

We inspired our neural network architecture on some of the MobileNetV4~\cite{MobinetV4} conceptual principles, with specific design variations. Our proposed architecture follows the MobileNet Inverted Residual Block approach, i.e., (down)scaled layers' dimensionality transitioning from low-dimensional space to high-dimensional space to low-dimensional space. 

\subsubsection{Spatial cross-antenna residual interference cancellation ($S$-Block in Fig.~\ref{fig:Hybrid_WAE_Diagram})} Different from MobileNet, to cancel the residual cross-stream interference at each temporal snapshot, spatial multi-user feature representations are jointly processed. The spatial features are processed through two fully connected dense layers with Mish activations $\psi(\cdot)$ and layer normalization $\Phi(\cdot)$:
\begin{equation}
	\mathbf{Z}_1 = \Phi\left(\psi\left(\mathbf{W}_{s1} \mathbf{X} + \mathbf{b}_{s1}\right)\right); \mathbf{Z}_2 = \Phi\left(\psi\left(\mathbf{W}_{s2} \mathbf{Z}_1 + \mathbf{b}_{s2}\right)\right)
\end{equation}
where the weight matrixes $\mathbf{W}_{s1}, \mathbf{W}_{s2} \in \mathbb{R}^{(3 N_{\text{tx}}) \times F_{\text{s}}}$ and $F_{\text{s}}$ denotes the dense layers dimension. We add Mish non-linear activation, i.e., $\psi(x) = x\tanh(\ln(1+e^x))$~\cite{Mish_2026}, because its smooth, continuously differentiable ($\mathcal{C}^\infty$) profile and slight negative activation overcome the hard, non-differentiable zero-thresholding of ReLU~\cite{Mish_2026}. At the end of the S-block, a residual skip connection is applied to preserve the signal representations, preventing gradient degradation during training, i.e., $\mathbf{Z}_{\text{r}} = \mathbf{Z}_1 + \mathbf{Z}_2$.

\subsubsection{Per-User-Stream temporal projection ($T$-Block in Fig.~\ref{fig:Hybrid_WAE_Diagram})} Following the initial spatial interference cancellation by the dense layers, each spatial stream will be processed over a few temporal symbols. For this, the joint spatial representations are mapped onto independent spatial tracks using an expanded feature projection layer over the temporal axis of the tensor, i.e., Spatio-Temporal Projection~(STP). The projection layer $Z_p$ will map $\mathbf{Z}_{\text{r}}$ to a combined and expanded dimensionality of $L \cdot F_{\text{t}}$, where $F_{\text{t}}$ will be the number of convolutional filters. Each of these filters will concurrently process a number of symbols equivalent to the filters' size.  

\begin{equation}
	\mathbf{Z}_{\text{c}} = \psi\left(\mathbf{W}_{\text{p}} \mathbf{Z}_{\text{r}} + \mathbf{b}_{\text{c}}\right) \in \mathbb{R}^{B \times N_{\text{sym}} \times (L \cdot F_t)}
\end{equation}
The output of the projected expanded layer is further reshaped and folded over the batch dimension, so that the  1D convolutional filters process each user stream in parallel and independently over the temporal axis of the tensor, i.e., $Z_t$. 

\begin{equation}
	\mathbf{Z}_t \in \mathbb{R}^{(B \cdot L) \times N_{\text{sym}} \times F_t}
\end{equation}

The convolutional filters are then defined using two 1D same-padded convolutional layers with Mish activations:

\begin{equation}
	\mathbf{Z}_{t1} = \psi\left(\text{Conv1D}_{k_1=5}\left(\mathbf{Z}_{t}\right)\right);\mathbf{Z}_{t2} = \psi\left(\text{Conv1D}_{k_2=3}\left(\mathbf{Z}_{t1}\right)\right)
\end{equation}
where the convolutional kernel sizes $k_1 = 5$ and $k_2 = 3$ aim to capture time-domain correlations over sliding multi-symbol windows across each independent user stream $L$. The reasoning is that, although the channel is memory-less the vertical interleaving of the same LDPC encoded stream across antennas inherently adds correlation.

\subsubsection{Demapping and Bit LLR Generation}
The fine-tuned modeling of the time-domain dependencies $\mathbf{Z}_{t2}$ is mapped directly to bit LLR logits using a last dense layer as a parametric function $ f_\Theta$. For this, we flatten the output dense layer $Z_f$:

\begin{equation}
	\label{eq:Zf_STp}
	\mathbf{Z}_f = \mathbf{W}_\text{o} \mathbf{Z}_{t2} + \mathbf{b}_\text{o} \in \mathbb{R}^{(B \cdot L) \times N_{\text{sym}} \times Q_m}
\end{equation}
Finally, the batch dimension is unfolded back to separate the individual user streams: 

\begin{equation}
	\label{eq:LLR_STp}
	\mathbf{LLR} = \text{Reshape}\left(\mathbf{Z}_f, \, [B, L, N_{\text{sym}}, Q_m]\right)
\end{equation}
The resulting $\mathbf{LLR}$ estimates are buffered until at least one full block length is demapped into $N$ bits. During inference time, batches of $B \times N$ bitwords are further forwarded to the primary inner decoder (i.e., LDPC decoder) for error correction and generating the estimated $K$ bits transmitted.

\subsection{End-to-End training through Binary Cross Entropy}

The spatially multiplexed end-to-end MIMO system can be parametrized at the transmitter side by means of the trainable complex constellation points $\mathbf{C} = \mathbf{C}_r + j \mathbf{C}_i \in \mathbb{C}^{2^{Q_m}}$, and at the receiver side by means of the neural decoder weights. To jointly train and optimize this system as a WAE, we need to define a bit-level proxy loss over the transmitted codeword bits. This is because, although the performance metric is the information bit error rate measured after the decoder ($\hat{\mathbf{b}}$), standard channel decoders are non-differentiable. Besides this problem, the forward error correction coding is also non-differentiable. Therefore, the reference for the loss function is taken on the LDPC coded bits $\mathbf{C} \in \{0, 1\}^{B \times N}$ prior to spatial multiplexing at the output of the transmitter. For the loss calculation, the LLR tensor is explicitly reshaped back to the original codeword dimensions to match the encoded bits $\mathbf{e} \in \{0, 1\}^{B \times N}$. The WAE is then trained using the Binary Cross-Entropy (BCE) with logits as the loss function:

\begin{multline}
	\label{eq:BCE}
	\mathcal{L}_{\text{BCE}}(\mathbf{e}, \mathbf{llr}) = -\frac{1}{BN} \sum_{i=1}^{BN} \Bigl[ e_i \ln \sigma(\text{llr}_i) \\
	+ (1 - e_i) \ln \bigl(1 - \sigma(\text{llr}_i)\bigr) \Bigr]
\end{multline}
where $\sigma(\cdot)$ is a sigmoid operator that converts raw network logits into soft posterior probabilities. Minimizing the BCE loss over the encoded bits will jointly optimize with the LDPC decoder the achievable block error rate over the batches of $B \times K$ information bitwords.

\subsection{BLER as a function of SNR}
As a benchmark, we consider an equivalent DNN-based end-to-end wireless autoencoder mapping either an M-NUC or M-QAM constellation. The wireless autoencoder's performance is evaluated by means of Monte Carlo simulations using Sionna’s BER utility library. This simulation framework allows us to estimate the $\text{BLER}$ as a function of the energy per information bit-to-noise power spectral density ratio ($E_b/N_0$) as well as the resulting system-level SNR. For each Monte Carlo iteration $j$, the block-level errors are evaluated and tracked by passing randomly generated data over the channel and retrieving it back from each receiver model under test:
\begin{equation}
	\text{BLER} = \frac{\sum_{j=1}^{J} \sum_{b=1}^{B} \mathbb{I}\!\left( \sum_{n=1}^{N} \left\vert{} b_{j,b,n} - \hat{b}_{j,b,n} \right\vert{} > 0 \right)}{D}
\end{equation}
where $b_{j,b,n}$ and $\hat{b}_{j,b,n}$ denote the transmitted and estimated bits, respectively, $\mathbb{I}(\cdot)$ represents the indicator function equal to $1$ if at least one bit in the block is received with an error, and $D$ is the total number of data blocks transmitted across all Monte Carlo iterations and batches.


The corresponding total SNR is scaled as a function of $E_b/N_0$ as:
\begin{equation}
	\mathrm{SNR} = \left( \frac{E_b}{N_0} \right)_{\mathrm{dB}} + 10 \log_{10} \left( \frac{K}{N} \cdot Q_m \right) \, [\text{dB}]
\end{equation}

\subsection{Computational Complexity and Energy Efficiency}
\label{subsec:EnergyMethods}

We account for the computational complexity based on the number of Multiply Accumulate Operations (MACs) of each block and its equivalent FLOPs. Although the operation count of the model can give an idea of its complexity, in a so diverse system design as the end-to-end wireless interface, it cannot always give the full picture. 
Therefore, for a fairer comparison, we 
designed an experimental testbench to measure the inherent power consumption linked to each algorithm, and compute the corresponding energy efficiency.

We  measured the GPU idle and algorithm execution power consumption (\textit{PC}) during thousands of iterations. Due to the non-deterministic nature of the operating system and the controlling drivers, several provisions are required to isolate the algorithm into a sandbox and minimize the influence of the operating system and driver controller, as well as external parameters like temperature.  To isolate the energy cost equivalent to the algorithm's arithmetic intensity, we measured first the idle power consumption floor of the GPU.  Over a calibration run consisting of $N_c$ idle samples, the total integrated idle energy ($E_{\text{idle}}$) is calculated as:
\begin{equation}
	E_{\text{idle}} = \sum_{j=2}^{N_c} \frac{PC_{j} + PC_{j-1}}{2} (t_j - t_{j-1})
\end{equation}

Because of the nondeterministic scheduling of the operating system, we cannot just subtract the difference to account for the net energy use, as there will be a mismatch between execution times. A correction is then applied to scale the net energy:
\begin{equation}
	E_{\text{net}} =
	\left[\sum_{i=2}^{N_s} \frac{PC_{i} + P_{i-1}}{2}\Delta t_i\right]
	-
	E_{idle}
	\cdot \frac{T_{\text{algo}}}{T_{\text{idle}}}
\end{equation}
where $T_{algo}$ and $T_{idle}$ are the execution times of the algorithm monitoring and calibration functions, respectively. The energy efficiency is then calculated considering how many bits were transmitted in data blocks without error:
\begin{equation}
	EE = \frac{D \cdot K \cdot (1 - \text{BLER})}{E_{\text{net}}} \, [\text{bits/J}]
\end{equation}

The measurements were performed on an NVIDIA RTX Pro 1000 Blackwell generation GPU. Some parameters on the GPU were fixed to limit their impact on the variability of the execution of the algorithm and power consumption measurements. First, the clock frequency of both the GPU core and memory is fixed to 1612~MHz and 9001~MHz, respectively. Also, allowing a too-long execution time $T_{algo}$ might cause a higher increase in temperature and the driver to reset the clocks for protection. Within an execution time of 2 minutes, this event never happened for an ambient temperature of $20^\circ\text{C}$. However, a small variation in the MACs' clock of 7~MHz occurred. The measurement sampling time was empirically determined within a range of 1ms and 500ms, with the lowest standard deviation at 200ms while keeping similar computational throughput. Controlling the thermal drift is also important for accuracy and reproducibility, as well as for keeping the initial conditions for each measurement. We set an initial GPU temperature of $42^\circ\text{C}$ and allow a cooling time between measurements of at least 2 min. During the cooling time, the GPU clocks are reset, and the temperature is monitored until it falls back and stabilizes at $42^\circ\text{C}$.

\section{Results and Analysis}
\label{sec:Results}
\subsection{Non-Trainable Hyperparameters}
To ensure a fair comparison, we select the minimum neural network size capable of separating all receiver constellation symbols, rather than minimizing BLER regardless of compute or energy cost. We evaluate this at a reference mid-sized block length ($K=1200$, $N=2400$) and vary hyperparameters across PHY-layer configurations only when the number of spatial streams or constellation size changes. Table~\ref{tab:phy_ml_hyperparameters} summarizes the non-trainable hyperparameters.

\begin{table}[htbp]
	\caption{Non-Trainable Hyperparameters}
	\label{tab:phy_ml_hyperparameters}
	\centering
	\small
	\scalebox{0.6}{
		\begin{tabular}{lccc}
			\hline
			\textbf{Hyperparameter / Parameter} & \textbf{Mid-Size PHY} & \textbf{Spatial Stream Scaling} & \textbf{Bitword Length Scaling} \\
			\hline
			\multicolumn{4}{l}{\textbf{PHY Layer Parameters}} \\
			Spatial Streams ($L$) & $4$ & $4, 8, 16$ & $8$ \\
			Information Bits ($K$) & $1200$ & $1200$ & $144 \text{ to } 3000$ \\
			Coded Block Length ($N$) & $2400$ & $2400$ & $288 \text{ to } 6000$ \\
			Constellation Size ($M$) & $16$ & $64$ & $64$ \\
			Bits per Symbol ($Q_m$) & $4$ & $6$ & $6$ \\
			\hline
			\multicolumn{4}{l}{\textbf{Spatio-Temporal Projected Network (WAE STP) Hyperparameters}} \\
			Spatial Dense Layer Size ($F_s$) & $16$ & $16,32,64$ & $32$ \\
			Projection Layer Size ($F_t \cdot L $) & $32$ & $64,128,256$ & $128$ \\
			Temporal Conv Filters ($F_t$) & $8$ & $16$ & $16$ \\
			Temporal Conv Kernels ($k_1, k_2$) & $5, 3$ & $5, 3$ & $5, 3$ \\
			Activation Function & Mish & Mish & Mish \\
			\hline
			\multicolumn{4}{l}{\textbf{Baseline Dense Neural Network (WAE DNN) Hyperparameters}} \\
			Spatial Dense Layer 1 Size ($F_{s1}$) & $32$ & $128$ & $128$ \\
			Spatial Dense Layer 2 Size ($F_{s2}$) & $32$ & $64$ & $64$ \\
			Activation Function & ReLU & ReLU & ReLU \\
			\hline
		\end{tabular}%
	}
\end{table}

\subsection{Mid-sized PHY Layer}
\begin{figure}[h]
	\centering
	\includegraphics[width=0.9\linewidth]{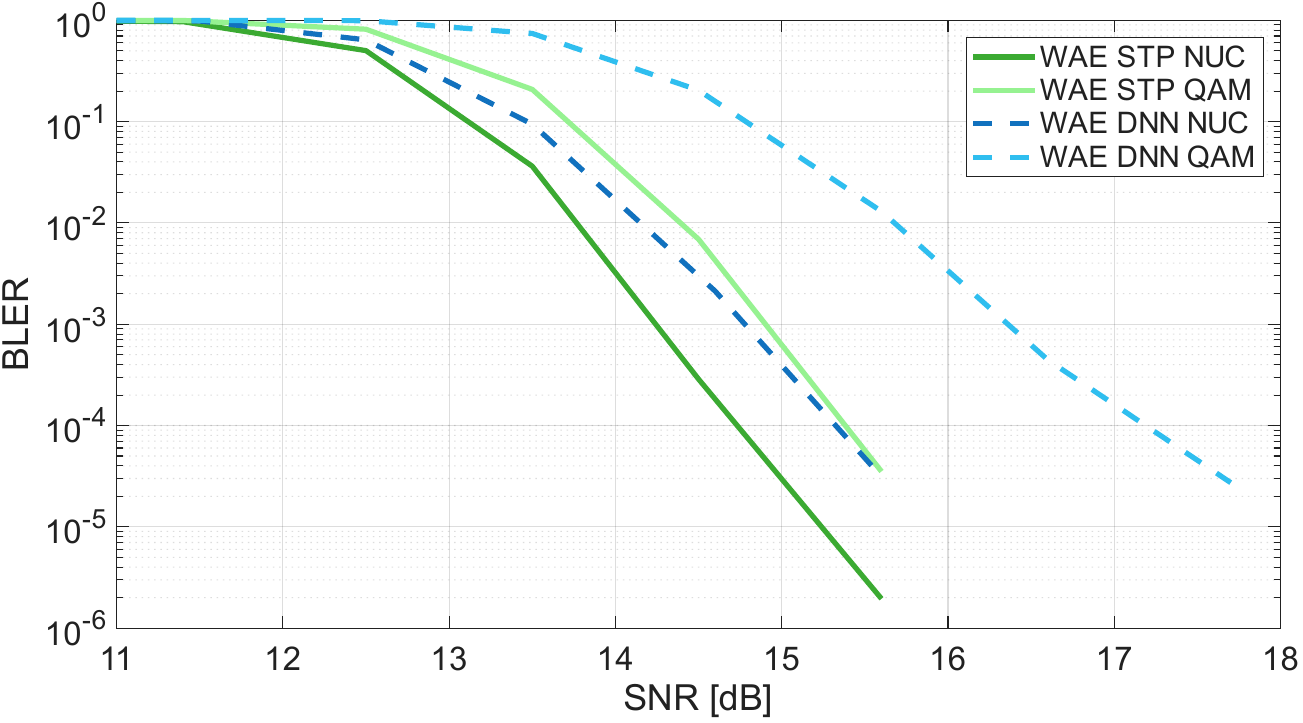}
	\caption{Block Error Rate of the Wireless Autoencoder for Middle-Size PHY.}
	\label{fig:BLER_L4_N2400_M16}
\end{figure}

We first compare the BLER of our proposed model to the benchmarking baselines for a small number of spatial streams (i.e., $L=4$) and considering moderate/mid-sized PHY parameters. Specifically, we first fix the encoded block length to $N=2400$ and an unencoded bitword of size $K=1200$, leading to a per-stream coding rate $ R=0.5$ and a constellation size of $M=16$ symbols. Fig.~\ref{fig:BLER_L4_N2400_M16} shows the BLER as a function of the SNR for each model.

The proposed WAE STP performs better than the WAE DNN for both modulation types, i.e., NUC and QAM. For the SNR threshold at which the WAE STP NUC achieves high reliability ($BLER \approx 10^{-3}$), the DNN-based autoencoder underperforms by a factor of more than 7. If we remove the gain of the NUC shaping from the autoencoders, there is a larger gain gap for the WAE DNN than for our designed WAE STP (i.e., 0.7 dB gap for the WAE STP and 1.5 dB for the WAE DNN). This is because the WAE STP not only adds constellation shaping gain but also higher robustness against implicitly correlated coded words across the spatial streams. The total gain of the WAE STP NUC for the ultrareliable threshold is approximately 2.2~dB. 

\begin{figure}[h]
	\centering
	\includegraphics[width=0.95\linewidth]{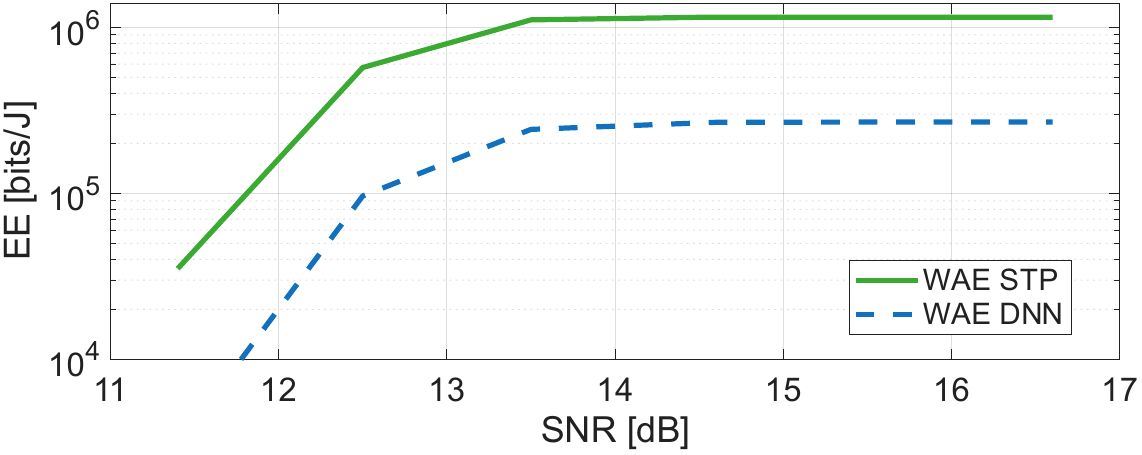}
	\caption{Energy Efficiency for Middle-Size PHY configuration.}
	\label{fig:EE_N2400_L4M16}
\end{figure}

In terms of computational complexity, the WAE STP model has a total number of operations of approximately 1.19 MFLOPs, while the WAE DNN requires approximately 22\% more. This is mainly because the input layer size of the WAE STP can be significantly reduced without a big impact on performance thanks to the following temporal processing layers. This difference in complexity becomes more significant in terms of energy efficiency in the GPU processing of the end-to-end models. Fig.~\ref{fig:EE_N2400_L4M16} shows the energy efficiency from the GPU measurements.   

We first evaluated the Batch size from the range $ B =\{32,64,128,256\} $. For the edge-grade GPU tested, the best performance was achieved for $B=64$. The WAE STP achieved a higher energy efficiency by a factor of 4.3 compared to the DNN-based baseline. This effect is not only caused by the difference in computational complexity. Compared to the DNN baseline, our proposed model allows for smaller-sized layers, which also helps maximize the GPU's MAC utilization in parallel threads. For instance, in this PHY configuration, our model uses dense layers of size $F_S = 16$ and only $F_t = 8$ convolutional filters, while for the DNN baseline the size of the dense layers is $F_S = 32$. This small difference reduced the DNN performance in terms of batch parallel processing to only 1859 iterations of $B=64$ on average, while our model was capable of performing 7 times more iterations on average. This led to a significant difference in the model's effective throughput for the same SNR threshold due to the computational constraints.  
 
\subsection{Model Scaling with the Number of Spatial Streams}
\begin{figure}[h]
	\centering
	\includegraphics[width=0.95\linewidth]{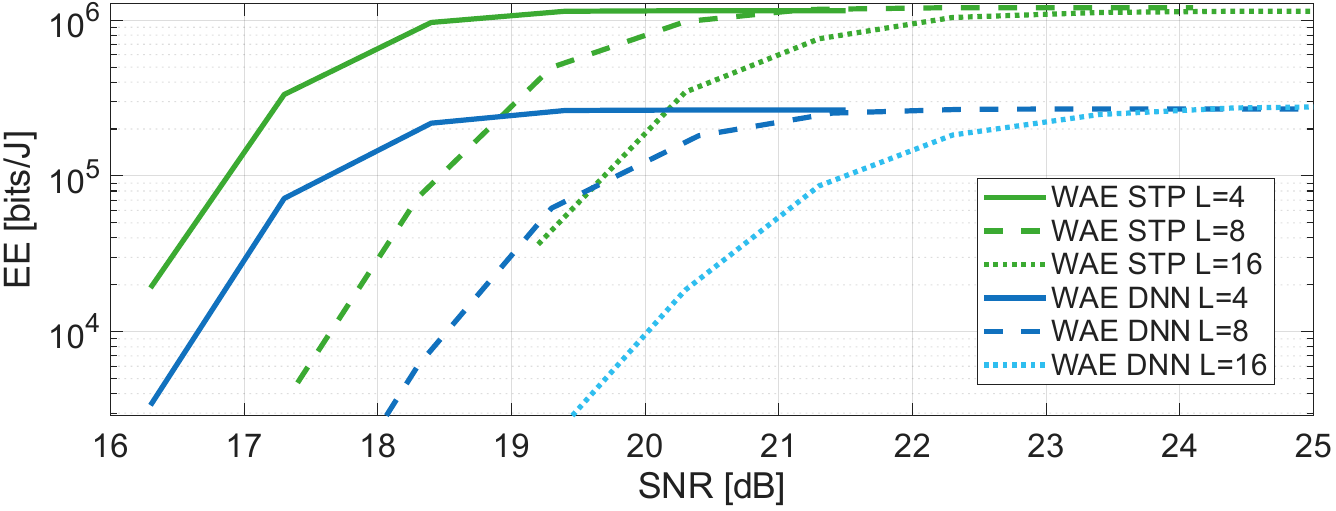}
	\caption{Energy Efficiency as a function of the number of spatial streams.}
	\label{fig:EEvsL}
\end{figure}

Fig.~\ref{fig:EEvsL} shows the scaling of the energy efficiency for different numbers of spatial streams. We analyze how each model scales for a higher constellation density (i.e., $M=64$) over a larger number of spatial streams (i.e., $L=\{4,8,16\}$).

At the computational saturation point, the designed WAE STP has a higher energy efficiency by a factor between 4.3 and 4.6. There is a combined effect of the BLER and the GPU peak efficiency in TOPS/W for a given level of parallelization. At lower SNRs, the energy efficiency is dominated by the BLER, before it saturates the GPU energy efficiency as a function of the number of spatial streams $L$. Under these computational constraints on edge devices, a larger number of antennas could reduce the required RF bandwidth due to a higher spectral efficiency, but it will not improve the computational energy efficiency. At lower SNRs, using fewer antennas can be even more advantageous in terms of computational energy efficiency, particularly for reconfigurable PHYs operating on power-constrained hardware.

\subsection{Model Scaling with the Block Length}

\begin{figure}[h]
	\centering
	\includegraphics[width=0.95\linewidth]{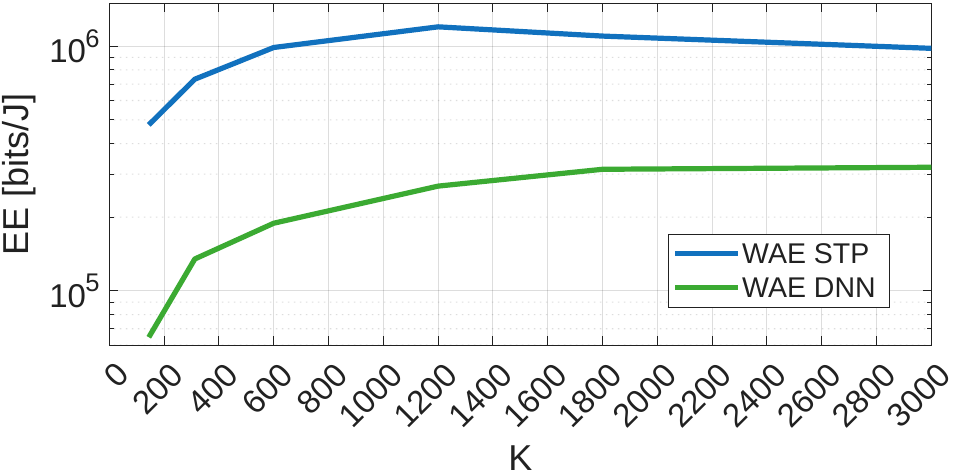}
	\caption{Energy Efficiency as a function of the uncoded bitword length K.}
	\label{fig:EEvsBlockLength}
\end{figure}

If we decrease $K$, then the peak parallelization can be reached for a higher $B$, but this requires buffering large batches of bitwords, which is counterintuitive to the reason why, for some applications, a smaller bitword is desired. For this reason, we keep the batch size constant at $B=64$. Here, we also set $L=8$, $M=64$, and the SNR at which the best-performing model achieves an ultrareliable link (i.e., $SNR \approx 22.2$~dB) as common references for the comparison. Fig.~\ref{fig:EEvsBlockLength} shows the variation in energy efficiency as a function of the bitword length for the aforementioned configuration.

For our WAE STP, the peak energy efficiency $(1.2 \cdot 10^6
)$ is reached for a bitword length of $K=1200$, being higher than for the WAE DNN by a factor of 4.5. From this bitword size onwards, the computational energy efficiency starts to decrease slightly. At this point, the LDPC complexity and the level of parallelization, as a function of the accelerator's MAC and memory bandwidth, become predominant. For $K=1200$, the complexity of the LDPC and the lack of further compute parallelization lean the balance towards a decline in peak compute performance per Joule consumed. For the DNN, there is still some margin on this trade-off as a function of $K,N$. However, notice that it never reaches the maximum energy efficiency of our proposed architecture. Even for $K=3000$, $N=6000$, our proposed design outperforms by a factor of 3.

\section{Conclusions}
In this paper, we presented a neural-based hybrid asymmetric MIMO wireless autoencoder. Our approach aims to reduce the gap between PHY virtualization and computational energy efficiency. It shifts computational complexity away from end-devices through direct constellation parameterization, without explicit MIMO precoding. The proposed design combines linear LMMSE spatial filtering with neural interference cancellation and demapping. It also introduces a spatio-temporally expanded projection architecture, consisting of an S-Block and a T-Block. These blocks decouple cross-antenna interference cancellation from temporal optimization. The proposed architecture adds robustness against implicit correlation from interleaving of the LDPC-encoded bitword across antennas. On edge-grade GPU hardware, the decoder achieves an energy efficiency exceeding $10^6$~bits/J, with a 3--4.6$\times$ improvement over DNN-based MIMO autoencoders.

\bibliographystyle{IEEEtran}
\bibliography{references}

@inproceedings{hesham_coding_2023,
	title = {Coding for the {Gaussian} {Channel} in the {Finite} {Blocklength} {Regime} {Using} a {CNN}-{Autoencoder}},
	doi = {10.1109/BlackSeaCom58138.2023.10299743},
	urldate = {2024-08-12},
	booktitle = {2023 {IEEE} {International} {Black} {Sea} {Conference} on {Communications} and {Networking} ({BlackSeaCom})},
	author = {Hesham, Nourhan and Bouzid, Mohamed and Abdel-Qader, Ahmad and Chaaban, Anas},
	month = jul,
	year = {2023},
	pages = {15--20},
}

@article{dorner_learning_2023,
	title = {Learning {Joint} {Detection}, {Equalization} and {Decoding} for {Short}-{Packet} {Communications}},
	volume = {71},
	issn = {1558-0857},
	doi = {10.1109/TCOMM.2022.3228648},
	number = {2},
	urldate = {2024-03-12},
	journal = {IEEE Transactions on Communications},
	author = {Dörner, Sebastian and Clausius, Jannis and Cammerer, Sebastian and Brink, Stephan ten},
	month = feb,
	year = {2023},
	pages = {837--850},
}

@article{ait_aoudia_waveform_2022,
	title = {Waveform {Learning} for {Next}-{Generation} {Wireless} {Communication} {Systems}},
	volume = {70},
	issn = {1558-0857},
	doi = {10.1109/TCOMM.2022.3164060},
	number = {6},
	journal = {IEEE Transactions on Communications},
	author = {Ait Aoudia, Fayçal and Hoydis, Jakob},
	month = jun,
	year = {2022},
	pages = {3804--3817},
}

@ARTICLE{hoydis_2021_6gainative,
  author={Hoydis, Jakob and Aoudia, Fayçal Ait and Valcarce, Alvaro and Viswanathan, Harish},
  journal={IEEE Communications Magazine}, 
  title={Toward a 6G AI-Native Air Interface}, 
  year={2021},
  volume={59},
  number={5},
  pages={76-81},
  doi={10.1109/MCOM.001.2001187}
}

@ARTICLE{HybridE2E_2026,
  author={Wei, Yuan and Chen, Chaoxu and Yao, Li and Zhang, Haoyu and Liu, Yinjun and Li, Fujie and Shen, Chao and Chi, Nan and Shi, Jianyang and Zhang, Junwen},
  journal={IEEE Transactions on Cognitive Communications and Networking}, 
  title={Advancing End-to-End Communication Systems With Physics-Aware Learning}, 
  year={2026},
  volume={12},
  number={},
  pages={10948-10964},
  doi={10.1109/TCCN.2026.3714071}
}

@software{sionna,
  title = {Sionna},
  author = {Hoydis, Jakob and Cammerer, Sebastian and {Ait Aoudia}, Fayçal and Nimier-David, Merlin and Maggi, Lorenzo and Marcus, Guillermo and Vem, Avinash and Keller, Alexander},
  note = {https://nvlabs.github.io/sionna/},
  version = {2.2.0}}

@techreport{etsi3gpp38212,
  author      = {{3GPP}},
  title       = {Multiplexing and channel coding},
  institution = {3rd Generation Partnership Project (3GPP)},
  type        = {Technical Specification (TS)},
  number      = {38.212},
  version     = {16.5.0},
  note        = {Release 16},
  url         = {https://3gpp.org},
}

@ARTICLE{TaskOriented_Comms,
  author={Xie, Songjie and Ma, Shuai and Ding, Ming and Shi, Yuanming and Tang, Mingjian and Wu, Youlong},
  journal={IEEE Journal on Selected Areas in Communications}, 
  title={Robust Information Bottleneck for Task-Oriented Communication With Digital Modulation}, 
  year={2023},
  volume={41},
  number={8},
  pages={2577-2591},
  doi={10.1109/JSAC.2023.3288252}
}

@ARTICLE{Song_2xMIMO_2022,
  author={Song, Jinxiang and Häger, Christian and Schröder, Jochen and O’Shea, Timothy J. and Agrell, Erik and Wymeersch, Henk},
  journal={IEEE Transactions on Wireless Communications}, 
  title={Benchmarking and Interpreting End-to-End Learning of MIMO and Multi-User Communication}, 
  year={2022},
  volume={21},
  number={9},
  pages={7287-7298},
  doi={10.1109/TWC.2022.3157467}
}

@ARTICLE{Velloso_2xMIMO_2026,
  author={Velloso, Eduardo N. and Silveira, Luiz F. Q. and Sousa, Elvino S.},
  journal={IEEE Access}, 
  title={End-to-End Autoencoder Design for MU-MIMO Systems With Low Pilot Overhead}, 
  year={2026},
  volume={14},
  number={},
  pages={18986-18993},
  doi={10.1109/ACCESS.2026.3659620}
}

@INPROCEEDINGS{Korpi_ICC_2026,
  author={Klus, Roman and Lampu, Vesa and Klus, Lucie and Korpi, Dani and Talvitie, Jukka and Raninen, Elias and Anttila, Lauri and Valkama, Mikko},
  booktitle={2026 IEEE International Conference on Communications Workshops (ICC Workshops)}, 
  title={DeepTRX: 6G Physical-Layer Neural Processing with Joint End-to-End Learning under Nonlinearities}, 
  year={2026},
  volume={},
  number={},
  pages={1-7},
  doi={10.1109/ICCWorkshops63917.2026.11586448}}

@ARTICLE{EqDeepRx_2026,
  author={Honkala, Mikko and Korpi, Dani and Raninen, Elias and Huttunen, Janne M. J.},
  journal={IEEE Transactions on Wireless Communications}, 
  title={EqDeepRx: Learning a Scalable and Interference Mitigating MIMO Receiver}, 
  year={2026},
  volume={},
  number={},
  pages={1-1},
  doi={10.1109/TWC.2026.3747193}
}

@InProceedings{MobinetV4,
author="Qin, Danfeng
and Leichner, Chas
and Delakis, Manolis
and Fornoni, Marco
and Luo, Shixin
and Yang, Fan
and Wang, Weijun
and Banbury, Colby
and Ye, Chengxi
and Akin, Berkin
and Aggarwal, Vaibhav
and Zhu, Tenghui
and Moro, Daniele
and Howard, Andrew",
editor="Leonardis, Ale{\v{s}}
and Ricci, Elisa
and Roth, Stefan
and Russakovsky, Olga
and Sattler, Torsten
and Varol, G{\"u}l",
title="MobileNetV4: Universal Models for the Mobile Ecosystem",
booktitle="Computer Vision -- ECCV 2024",
year="2025",
publisher="Springer Nature Switzerland",
address="Cham",
pages="78--96",
isbn="978-3-031-73661-2"
}

@INPROCEEDINGS{Mish_2026,
  author={Liu, Junting and Shu, Mingyu and Liu, Qiang},
  booktitle={2026 IEEE International Symposium on Circuits and Systems (ISCAS)}, 
  title={Efficient Hardware-Oriented Approximation of Mish Activation Function and Its Realization}, 
  year={2026},
  volume={},
  number={},
  pages={2188-2192},
  doi={10.1109/ISCAS66217.2026.11562498}}

\vspace{12pt}

\end{document}